\documentclass[aps,prl,twocolumn,groupedaddress,showpacs]{revtex4}

\usepackage{graphicx}

\begin{document}

\preprint{PRL/Chen, \textit{et. al.}}

\title{Adaptive control of CO$_2$ bending vibration: deciphering field-system dynamics}

\author{G.-Y. Chen,$^1$ Z. W. Wang,$^2$ and W. T. Hill, III$^{1,2,3}$}

\affiliation{$^1$Department of Physics, $^2$Institute for Physical Science and Technology
and $^3$Joint Quantum Institute \\ University of Maryland, College Park, Maryland 20742}

\email{wth@umd.edu}

\date{\today}

\begin{abstract}

We combined adaptive closed-loop optimization, phase-shaping with a restricted search
space and imaging to control dynamics and decipher the optimal pulse. The approach was
applied to controlling the amplitude of CO$_2$ bending vibration during strong-field
Coulomb explosion. The search space was constrained by expressing the spectral phase as a
Taylor series, which generated pulses with characteristics commensurate with the natural
physical features of this problem. Optimal pulses were obtained that enhanced bending by
up to 56\% relative to what is observed with comparably intense, transform limited
pulses. We show that (1) this judicious choice of a reduced parameter set made unwrapping
the dynamics more transparent and (2) the enhancement is consistent with field-induced
structural changes to a bent excited state of CO$_2^{2+}$, which theoretical simulations
have identified as the state from which the explosion originates.

\end{abstract}


\maketitle

The confluence of intense laser fields of ultra-wide spectral bandwidth, pulse
shaping \cite{Weiner00} and adaptive feedback \cite{Levis01} provides a unique
opportunity to explore the manipulation and engineering of many-particle
dynamics at the quantum level. From two-particle collective motion to the
complexities of biological interactions, the potential to guide an arbitrary
system with tailored pulses in a predetermined way has opened new vistas for
control via light-matter interaction. While much progress has been made towards
developing schemes for realizing control -- achieving a specific goal
\cite{Judson92,Warren93} -- an ability to relate the complex field patterns of
the control pulse to a sequence of steps along and/or between potential
surfaces describing specific states of the system remains illusive, in general.
Two of the primary impediments inhibiting advance are the vast number of (1)
degrees of freedom available to the system under control and (2) phase and/or
amplitude parameters influencing the pulse that must be set. To appreciate the
magnitude of the latter, for phase only shaping with just 128 spectral
divisions each with say 700 possible steps between 0 and $2 \pi$, there are
$\sim 1.5 \times 10^{364}$ permutations. Suggestions for reducing the set have
ranged from limiting the choices to binary values (0 or $\pi$) to imposing a
specific functional form (see \cite{Lozovoy2008, Efimov1998a, Efimov1998b}, for example).  In addition to
reducing the search space, the "right" choice -- one natural to the physics of
the problem -- can be quite powerful, facilitating efficient genetic (GA) or
evolutionary algorithm searches and, more importantly, enabling solutions to be
deciphered physically. In this paper we demonstrate the power of a reduced set
pulse shaping in a GA-mediated closed-loop control experiment designed to
enhance bending of CO$_2$ during strong-field Coulomb explosion at 800 nm.
Specifically, we enhanced bending for the symmetric 6-electron channel,
CO$_2^{6+} \rightarrow $ O$^{2+}$ + C$^{2+}$ + O$^{2+}$, in which $p_{||}=0$
for C$^{2+}$, $p_{||}$ is equal and opposite for O$^{2+}$ and $p_\perp$ for
C$^{2+}$ is equal and opposite to the sum of $p_\perp$ for O$^{2+}$, where $p_{||,\perp}$
is the momentum relative to the $\vec{E}$ field of the
light.

We chose CO$_2$, which is linear in its ground state ($\Delta \theta_b = 0$, the
deviation away from $180^\circ$), as our target because it is a nontrivial molecular
system exhibiting modes common in larger systems -- stretching and bending -- and it is
well known to bend significantly ($\Delta \theta_b > 0$) during strong-field induced
Coulomb explosion \cite{Cornaggia96a,Hishikawa99a,Zhao03}. The large bending has also be
the subject of theoretical studies, which suggest a physical explanation for the
distortion \cite{Kono2001,Sato03,Kono2006}. The CO$_2$ response to strong fields allows
us to rely on Coulomb explosion imaging \cite{Zhao01} to probe the connection between the
optimal field solutions and the dynamics, where the strong field pulse is the
protagonist, staring the control agent and the detector.  Our goal was to verify the
theoretical model experimentally by exploiting it to enhance the distortion.

\begin{figure}[t]
    \includegraphics[scale=0.3, angle=0]{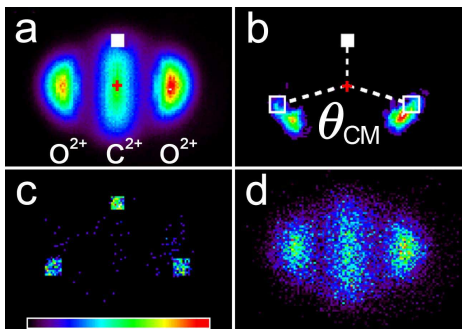}
    \caption{(Color online.) Images of the symmetric 6-electron Coulomb explosion
    channel of
    CO$_2^{6+}$ induced by a shaped pulse (top row, Fig.\ \ref{fig:GA}):
    (a) digital composite with 2,500,000 shots; (b) selective average \cite{Zhao03},
    all frames in (a) with C$^{2+}$ landing inside the filled square; (c) triple-coincidence \cite{zhao05}, all frames in (b) with ions landing
    simultaneously in pre-selected areas obeying momentum conservation (filled and hollow squares) and (d) analog composite of 15,000 shots.  In all panels, $\vec{E}$ is horizontal and the center is the center of mass of the explosion.  Panel (b) defines the far-field angle between the two
    O$^{2+}$ momenta, $\theta_{CM}$, 147$^{\circ}$ in this case.
    }\label{fig:images}
\end{figure}

In our experiment, 50 fs transform limited (TL) pulses from a Ti:sapphire laser
system were shaped by a 128 element phase only liquid crystal spatial light
modulator (SLM). Pulses were focused with a spherical mirror (f.l.\ 75 mm) to
an $\sim 8$ $\mu$m waist in a chamber containing $\sim 5 \times 10^{-8}$ Torr
of CO$_{2}$.  The ions generated by the Coulomb explosion were detected with
our $4 \pi$ image spectrometer \cite{Zhu97,Zhao02}.  Microchannel plates backed
by a phosphor generate visible images of 2-dimensional projections of
3-dimensional momentum distributions of the ions. Images were collected with an
analog camera capable of streaming frames to disk at 15 Hz or a fast-frame
digital camera with a frame-storage rate of $730$ Hz (see in Fig.\
\ref{fig:images}).  The laser rep rate was set to match the digital camera rate
to ensure one shot per frame. Typically, the digital (analog) composite images
contained $\sim 2,500,000$ ($\sim 15,000$) laser shots.

\begin{figure}[t]
    \includegraphics[scale=0.65, angle=0]{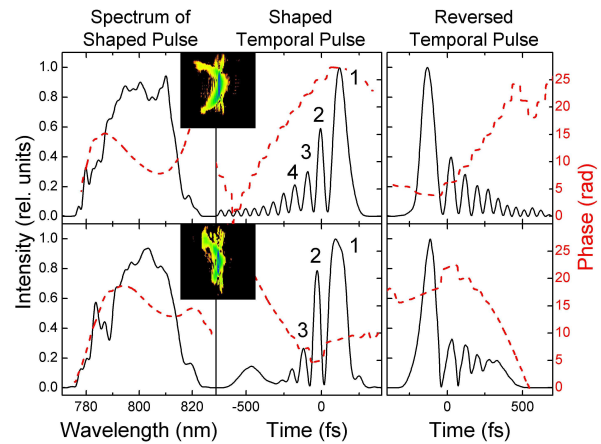}
    \caption{(Color online.) Reconstructed distributions from FROG traces (insets)
    for the two GA solutions (left and middle panels) and their phase reversals (right
    panel) corresponding to the filled and hollow diamonds (upper row) and triangles
    (lower row) in Fig.\ \ref{fig:I-dist}:  the solid black curves are the spectra
    and the temporal intensities (left axes) and the dashed red curves are the
    spectral and temporal phases (right axes).  The reversed cases have the same spectra.
    The vertical axes of the FROG traces (insets) are the
    wavelength ranging from 768 nm to 837 nm and horizontal the delay time from
    -1666 fs to 1666 fs. The parameters for the labeled peaks are given in
    Table\ \ref{Tab:pulse}. Coefficients for the $n=2,..,5$ spectral
    phase terms are respectively:  800 fs$^2$, -$1.98 \times 10^5$ fs$^3$,
    -$3.47 \times 10^6$ fs$^4$ and -$5.30 \times 10^8$ fs$^5$ (upper left panel) and
    -2954 fs$^2$, -$5.86 \times 10^5$ fs$^3$, $1.74 \times 10^5$  fs$^4$ and $1.11 \times
    10^9$ fs$^5$
    (lower left panel).  For the reverse cases, the sign is reversed in all terms.
    }\label{fig:GA}
\end{figure}

Searches were performed on analog images using the length of the C$^{2+}$ lobe as the
fitness parameter. This was possible since C$^{2+}$ is constrained to motion $\perp$ to
$\vec{E}$ so larger $\Delta \theta_b$ produces longer lobes. The search space was reduced
by restricting the phase mask such that the spectral phase was expressed as a
5$^{\mathrm{th}}$-order Taylor expansion \cite{Efimov1998a,Efimov1998b}, $\varphi(\omega)=\sum^5_{n=0}
\varphi_{n}(\omega_0)(\omega-\omega_{0})^n/n!$, where $\varphi_{n}(\omega_0) = \partial^n
\varphi(\omega)/\partial \omega^n |_{\omega_{0}} $.  We ignored the first two terms of
the series because they determine the ``absolute'' phase and group delay, neither of
which is important here. Thus, the GA only varied four parameters, creating a phase mask
by adjusting the pixels of the SLM collectively.  It is clear from Fig.\ \ref{fig:GA}
that solutions are pulse trains that can be described in terms of even and odd spectral
phase orders, giving us important insight into the pulse and, the dynamics it induces, as
we discuss below.  Characterization of the solution is critical for this study. A
transient-grating FROG (Frequency-Resolved Optical Gating \cite{Trebino2002}) was our
primary tool for this purpose; Fig.\ \ref{fig:GA} shows the FROG traces for the GA
solutions as well. Consistency checks and calibration of the FROG phase retrieval were
verified with a SPIDER (Spectral Phase Interferometry for Direct Electric-field
Reconstruction \cite{SPIDER}) and a SEA-TADPOLE (Spatial Encoded Arrangement for Temporal
Analysis by Dispersing a Pair of Light E-fields \cite{TrebinoSEATADPOLE}) arrangements for
several different pulse shapes.

\begin{figure}[t]
    \includegraphics[scale=0.5, angle=0]{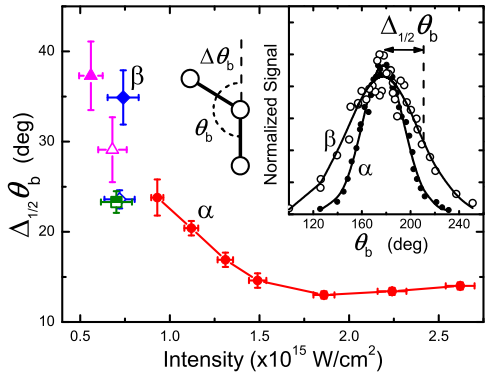}
    \caption{(Color online.) Bond angle distributions obtained from a triple-coincidence
    measurement
    and their Gaussian fits (upper right, for points $\alpha$ and $\beta$)
    and $\Delta_{1/2}\theta_b$ vs.\ $I$ for 50 fs transform limited
    pulses (filled red circles), 100 fs transform limited pulse (half-filled green
    square) and four shaped pulses, the two GA solutions in the upper and lower panels
    of Fig.~\ref{fig:GA} (filled diamond and triangle respectively) and their reversals
    (hollow diamond and triangle) as discussed in the text.  The errors in intensity
    reflect the fluctuations (standard deviation) in the power and pulse-width measurements,
    while those in
    $\Delta_{1/2}\theta_b$ are due to the standard deviations in the fit parameters.
    The cartoon defines the angles.
    }\label{fig:I-dist}
\end{figure}

Once an optimal pulse was found, statistical correlation techniques -- \textit{image
labeling} \cite{Zhao01}, \textit{selective averaging} \cite{Zhao03}, and
\textit{coincidence imaging} \cite{zhao05} -- were run on digital images to identify
collision partners and to measure relative yields vs.\ $\Delta \theta_b$. Results are
displayed in Fig.\ \ref{fig:I-dist}.  To determine $\Delta \theta_b$, we measured the
far-field center of mass angle, $\theta_{CM}$, defined in Fig.\ \ref{fig:images}b, from
the locations of the correlated O$^{2+}$ obeying momentum conservation (the hollow
squares in Fig.\ \ref{fig:images}b). The bond angle, $\theta_b = 180^\circ - \Delta
\theta_b$, at the time of explosion was determined from $\theta_{CM}$ numerically from
the equations of motion assuming a pure Coulomb explosion. We point out that
$\theta_b(\theta_{CM})$ depends weakly on the value of $R_{ex}$ \cite{Zhao03}, the
explosion bond length, and the charge, $qe$, deviates by less than 1\% for $R_{ex}$
in the 2.3 to 4.1 atomic units (a.u.) range and $q$ between 1 to 2. Thus, any residual
bonding and/or variation in $R_{ex}$ will have little affect on the angles we report.
Distributions similar to that shown in the inset of Fig.\ \ref{fig:I-dist} for different
angles were obtained by selecting different C$^{2+}$ ions along the central lobe.

Our investigation was composed of three experiments to probe the explosion response to
the field. To isolate the shaped-pulse effects from changes induced by merely varying the
intensity or the duration of the pulse, we performed two experiments with TL pulses.
First, we measured the $\theta_b$ distribution vs.\ $I$ for 50 fs pulses, the solid curve
in Fig.\ \ref{fig:I-dist}. The half width of the distribution, $\Delta_{1/2}\,\theta_b$,
decreases monotonically with increasing $I$, $\sim 14^\circ$ when $I \gtrsim 1.8 \times
10^{15}$ W/cm$^2$, increasing to $\sim 24^\circ$ at $\sim 9 \times 10^{14}$ W/cm$^2$, the
lowest intensity at which we could analyze images due to lack of signal strength, a 71\%
enhancement. Second, we measured the distribution for a 100 fs TL pulse at an intensity
commensurate with the shaped-pulse intensity ($\sim 7 \times 10^{14}$ W/cm$^2$), the half
filled square in Fig.\ \ref{fig:I-dist}. While the signal yield was considerably stronger
with the longer pulse, the bending response did not improve over the best 50 fs pulse
response.  We note that $\Delta_{1/2} \theta_b$ decreased as $I$ increased for 100 fs
pulses as well. Finally, we measured the distribution for two GA solutions, the filled
triangle and diamond in Fig.\ \ref{fig:I-dist}. We plotted the GA solution at the
intensity of the largest peak in the train. From the table of pulse train parameters
(Table \ref{Tab:pulse}) the width of the largest peak is 100 (128) fs for the upper
(lower) solution in Fig.~\ref{fig:GA}. Clearly, the pulse train induces considerably more
bending than does a TL pulse, providing an additional enhancement of 46 - 56\% (compare
the 100 fs TL pulse response with that of the two GA solutions).

We begin our discussion of the results by looking at the system response to TL fields.
The sensitivity to $I$ depends on several factors.  First, it is well known that enhanced
ionization \cite{Seideman95,Chelkowski95b,Zuo95}, mediated by over-the-barrier
ionization, is the principal ionization mechanism responsible for Coulomb explosion for
pulses longer than 30 fs \cite{Alnaser04,Alnaser04b}. The critical bond length, $R_C$,
where the barrier is lowest, is system dependent, going as $\sim 2.5/I_P$ a.u.\ for
linear tiatomics \cite{Yu98} where $I_P$ is the atomic ionization potential.  For CO$_2$,
$R_C$ is in the 3.5 to 5 a.u.\ range; we measured $R_C \simeq 4$ a.u.\ \cite{Zhao03}.
Second, theoretical simulations suggest the first two electrons are removed from CO$_2$
before the explosion \cite{Kono2001,Sato03}, which is consistent with our earlier
measurements \cite{Zhao03}. Third, the same simulation shows CO$_2^{2+}$ is promoted to a
bent excited state prior to Coulomb explosion. The theory focused on the first two
adiabatic states for CO$_2^{m+}$ ($m=0,1,2$), $|1\rangle_{m+}$ and $|2\rangle_{m+}$, and
found that $|2\rangle_{2+}$ was bent (Fig.\ 7 in \cite{Sato03}) while the other five were
not. The resonant frequency for the CO$_2^{2+}$ transition, $|1\rangle_{2+} \rightarrow
|2\rangle_{2+}$, was calculated to close to 800 nm near $R_C$ \cite{Sato03}. When a
system makes a transition from a linear state to a bent state it will be end up in a high
vibrational level and vibrate with larger amplitude. Fourth, the field-dressed states are
flatter in the $R$ coordinate than the field-free states (see Figs.\ 3, 6 and 9 in
\cite{Kono2001}) and the dressed $|2\rangle_{2+}$ state is less bent, $\Delta \theta_b \sim
15^\circ$, than it is field-free, $\Delta \theta_b \sim 60^\circ$ (compare Fig.\ 7 in
\cite{Sato03} with Fig.\ 17 in \cite{Kono2006}). Flatter potentials allow the system to
stretch further making it easier to reach $R_C$. Consequently, theory suggests that
Coulomb explosions originating from $|2\rangle_{2+}$ will exhibit larger bending
amplitude than those originating from $|1\rangle_{2+}$. We expect less bending from
$|2\rangle_{2+}$ when dressed with a stronger field because the $|1\rangle_{2+}
\rightarrow |2\rangle_{2+}$ transition could promote population to lower vibrational
states in $|2\rangle_{2+}$. This partially explains our observations. We point out,
however, that with the observed increase in the bending for decreasing $I$ (Fig.\
\ref{fig:I-dist}), is a concomitant, monotonic decrease in kinetic energy release during
the explosion, indicating $R_{ex} \propto 1/I$. This well known saturation is a result of
the threshold $I$ for over-the-barrier ionization increasing as $R$ decreases. For strong
enough fields, over-the-barrier ionization and the explosion occur at $R_{ex} < R_C$.
Thus, near $R_C$ Coulomb explosion is induced at lower $I$, which again leads a larger bending for smaller $I$.

\begin{table}[t] %add [H] placement to break table across pages
\caption{Pulse train parameters for the GA solutions in Fig.\ \ref{fig:GA} with the
(top/bottom) values corresponding to the (upper/lower) GA solutions and the column
headings correlate with the peaks in Fig.\ \ref{fig:GA}: $\tau_i$ (fs) is the pulse
duration; $\Delta t_{ij}$ (fs) is the period since the previous peaks; $\varepsilon_i$ is
the energy of the $i^{th}$ peak relative to the total energy in the shaped pulse; and
$I_i$ ($\times 10^{14}$ W/cm$^2$) is the intensity of the $i^{th}$ peak.   The
corresponding values for the largest peak of the reversal for each GA solution are:
100/104 for $\tau_i$, 0.62/0.59 for $\varepsilon_i$ and 7.2/6.8 for $I_i$.}
%\begin{footnotesize}
%\begin{ruledtabular}
\begin{tabular}{c@{~~~~}c@{~~~~}c@{~~~~}c@{~~~~}c}
\hline \hline
  & 1 & 2 & 3 & 4  \\
\hline
 $\tau_i$ & 100/128 & 47/52 & 46/48 & 44/N.A. \\
 $\Delta t_{ij}$ & 120/127 & 87/87 & 87/N.A. & N.A./N.A.  \\
 $\varepsilon_i$ & 0.63/0.57 & 0.17/0.19 & 0.08/0.06 & 0.06/N.A. \\
 $I_i$ & 7.5/5.6 & 4.2/4/3 & 2.1/1.5 & 1.6/N.A. \\
\hline & & & & \\
\end{tabular}\label{Tab:pulse}
%\end{ruledtabular}
%\end{footnotesize}
\end{table}

Turning to the GA solutions (Fig.\ \ref{fig:GA}), we notice that while the two pulse
trains are not identical (see Table \ref{Tab:pulse} for a summary of their
characteristics), they do share two important features: (1) the last peak is larger and
wider than the earlier peaks -- a pulse width $\gtrsim 100$ fs compared with $\sim 50$ fs
for early pulses -- and (2) the spectral phase is dominated by a negative odd (largely
third) order (see the caption of Fig.\ \ref{fig:GA} for the $\varphi_n$ values). A pulse
train preceding a larger peak is characteristic of a negative odd order spectral phase.
In fact, all solutions shared these two features.  All solutions showed bending
enhancement over TL pulses, albeit, with some inducing significantly more bending than
others.  One advantage of this reduced parameter set is that we can change the sign of or
turn off specific phase terms.  To verify that the negative odd order spectral phase was
in fact important, we reversed the sign of the $\varphi_n$ terms to produce a pulse train
with the first peak being largest (right panels in Fig.\ \ref{fig:GA}).  The result was
the enhancement virtually disappeared as shown by the hollow diamond and triangle in
Fig.\ \ref{fig:I-dist}. The most likely reason the train of pulses enhances bending
vibration is a combination of the fact that the $|2\rangle_{2+}$ state is populated by
the early peaks more completely than with a single pulse and the field oscillates off and
back on.  As we pointed out earlier, atomic ions were essentially immeasurable for $I
\lesssim 7 \times 10^{14}$ W/cm$^2$.  Even though Coulomb explosions were insignificant,
this intensity was sufficient to produce both CO$_2^+$ and CO$_2^{2+}$ ions.  So, each of
the early peaks help to populate the $|2\rangle_{2+}$ state. Simulations show that
subsequent peaks do not transfer population back to the $|1\rangle_{2+}$ state
\cite{Kono:2008}. Between pulses, when the field is off, the state relaxes to its more
bent field-free condition causing CO$_2^{2+}$ to vibrate more. It is also interesting to
note that the period between pulses is near the bending vibrational frequency of the
ground state. Thus, the field could also provide a periodic kick, further enhancing the
vibration. It is important to see that the GA enhancement does not come from a lowering
of $I$. The 100 fs TL and the two reversal solutions all show significantly less bending
than the GA solutions but have $I$ about the same as that of the diamond solution. The
kinetic energy release for the filled triangle and filled diamond are identical. The fact
that the kinetic energy release for the GA solutions was identical suggest that any
enhancement due to intensity and bond length at the time of explosion is probably
minimal. We then conclude that the pulse train with a negative third (odd) order chirp
plays a primary roll in enhancing the bending amplitude.

The results presented here coupled with our earlier studies \cite{Zhao03} are consistent
with the theoretical model of the CO$_2$ explosion originating from a high vibrational
level of a bent state in CO$_2^{2+}$.  For TL pulses, the bending amplitude decreases
with increasing $I$, in response to a straightening of the bent state for larger $I$. A
shaped pulse composed of a pulse train provides additional bending enhancement.  The
dressed-bent state is allowed to relax (become more bent) between peaks within the train
amplifying the vibration, with the early peaks populating the state and final peak
inducing the explosion. A judicious choice for the search space proved to be powerful in
deciphering the optimal pulse. Vibration was the natural physical feature for this
problem and restricting the phase mask to producing pulses forced the optimal pulse to
reflect the primary dynamics.

We thank Prof. H. Kono for helpful discussions, Dr. G. M. Menkir for helpful
discussions and technical support in the early stages of this work and S.
Iacangelo (REU student), T. Avasthi and J. Sun for technical support.  This
work was supported by NSF grant PHY0555636.

\bibliography{shaped}

\clearpage

\end{document}